\documentclass[twocolumn,showpacs,amsmath,amssymb,superscriptaddress]{revtex4}
\usepackage{graphicx}
\usepackage{dcolumn}
\usepackage{bm}

\begin{document}

\title{Symmetry energy properties of neutron-rich nuclei from the coherent density
fluctuation model applied to nuclear matter calculations with Bonn potentials}

\author{I. C. Danchev}
\affiliation{Department of Physical and Mathematical Sciences, School of Arts and Sciences,
University of Mount Olive, 652 R.B. Butler Dr., Mount Olive, NC 28365, USA}

\author{A. N. Antonov}
\affiliation{Institute for Nuclear Research and Nuclear Energy,
Bulgarian Academy of Sciences, Sofia 1784, Bulgaria}

\author{D. N. Kadrev}
\affiliation{Institute for Nuclear Research and Nuclear Energy,
Bulgarian Academy of Sciences, Sofia 1784, Bulgaria}

\author{M. K. Gaidarov}
\affiliation{Institute for Nuclear Research and Nuclear Energy,
Bulgarian Academy of Sciences, Sofia 1784, Bulgaria}

\author{P. Sarriguren}
\affiliation{Instituto de Estructura de la Materia, IEM-CSIC, Serrano 123,
E-28006 Madrid, Spain}

\author{E. Moya de Guerra}
\affiliation{Departamento de Estructura de la Materia, F\'\i sica T\'ermica y
Electr\'onica, and IPARCOS, Facultad de Ciencias F\'\i sicas, Universidad
Complutense de Madrid, Madrid E-28040, Spain}

\begin{abstract}
We derive the values of nuclear symmetry energy, its components, as well as
pressure in finite nuclei at saturation density from their corresponding values
in nuclear matter obtained in non-relativistic Brueckner-Hartree-Fock
calculations with the realistic Bonn B and Bonn CD potentials using the coherent
density fluctuation model in the framework of a self-consistent
Skyrme-Hartree-Fock plus BCS method. We focus on three isotopic chains of
spherical nuclei (Ni, Sn, and Pb) and compare our results with those obtained
with an effective Brueckner density-dependent potential. The role of the
three-body forces on the considered quantities is also studied and discussed.
\end{abstract}


\maketitle

\section{Introduction \label{sec:1}}

The basic properties of many-body quantum-mechanical systems with strongly
interacting particles primarily depend on the equation of state (EOS). In
particular, this concerns symmetric and asymmetric nuclear matter including
astrophysical objects such as neutron stars and the gravitational wave events
caused by binary black hole and binary neutron star mergers (see, e.g.,
Ref.~\cite{Bombaci2018}). A very important theoretical problem is to determine
the EOS from underlying nuclear interactions using two-, three-, and many-body
nucleon interactions derived in various theoretical approaches. It is necessary
in many cases to calculate the EOS to extreme conditions of high density and
high neutron-proton asymmetry. This inevitably leads to the necessity to have a
knowledge of the nuclear symmetry energy $S^{NM}(\rho)$ as one of the most
important quantities that determines the structure and the pressure of the
nuclear matter (NM). The $S^{NM}(\rho)$ is the difference between the binding
energy per particle in pure neutron matter and in symmetric nuclear matter. Many
calculations of the EOS and $S^{NM}(\rho)$ have been performed using various
realistic models of the nucleon-nucleon interactions with parameters adjusted so
as to reproduce the observed nucleon-nucleon ($NN$) scattering phase shifts.
Here we note (see \cite{Engvik1997}) the variational calculations in
Ref.~\cite{Wiringa1988} with Argonne $V_{14}$ potentials \cite{Wiringa1984} and
the earlier Brueckner-Hartree-Fock (BHF) calculations using one-boson-exchange
potential \cite{Engvik1996,Baldo1997} with Reid \cite{Reid1968} and Paris
potentials \cite{Lacombe1980}. It has been concluded in Ref.~\cite{Engvik1997}
that the differences in the predictions of the $S^{NM}(\rho)$ at high densities
are not caused by the many-body method used, but rather by the various models
used for the $NN$ interactions.

In what follows we concentrate on the non-relativistic many-body BHF
calculations in NM with various potentials. It was initiated in the sixties by
Brueckner \cite{Brueckner-Levinson1955,Brueckner1955} and Bethe
\cite{Bethe1956}, and later by Haftel and Tabakin \cite{Haftel-Tabakin1970} and
Erkelenz {\it et al.}
\cite{Erkelenz-Alzetta-Holinde1971,Erkelenz-Holinde-Bleuler1971} with the
objective to derive the saturation properties of nuclear matter from first
principles and obtain the experimentally observed nuclear binding energy.
Self-consistent calculations of the density dependence of the nuclear symmetry
energy were carried by Brueckner, Coon, and Dabrowski in
Ref.~\cite{Brueckner-Coon-Dabrowski1968}.

Substantial progress in the study of the equation of state of nuclear matter was
achieved in the nineties with relativistic extensions of that approach and
employment of one-boson-exchange nucleon-nucleon potentials
\cite{Machleidt_ANP1989,Machleidt2001,Machleidt_CompNuc2}. The investigations of
$\beta$ stability of nuclear matter \cite{Prakash1994} at high densities brought
to center stage the need of accurate knowledge of the density dependence of the
nuclear symmetry energy and many-body calculations in both non-relativistic
\cite{Engvik1997} and relativistic \cite{CHLee1998} framework were carried out
using modern realistic potentials.

A comparative study of different approaches for the properties of asymmetric
nuclear matter has been presented in Refs.~\cite{Gogelein2009,Hassaneen2014}.
These various approximation schemes are shown to lead to rather similar
predictions for the energy per nucleon of symmetric and asymmetric nuclear
matter at high densities and large proton-neutron asymmetries.

The study of finite nuclei has further significance for the development of
nuclear density functionals. Very recently, Shen {\it et al.} have developed and
realized an application of relativistic BHF theory for finite nuclear systems
\cite{Shen2019}. They showed that further improvements in building {\it ab
initio} relativistic energy functionals are needed (inclusion of higher orders
of the hole-line expansion, more precise relativistic $NN$ interactions, efforts
to develop relativistic chiral $NN$ interactions) to go, for instance, to
heavier systems.

The main task of our work is to employ the modern realistic Bonn B
and Bonn CD potentials on the basis of the non-relativistic
approach used by Engvik {\it et al.} \cite{Engvik1997}. First, we
obtain the density dependence of the NM symmetry energy for a
range of densities relevant to the present study and then we use
the coherent density fluctuation model (CDFM) \cite{Antonov80,AHP}
to extract the symmetry energy and pressure in three isotopic
chains of (finite) spherical nuclei with input nuclear densities
derived in self-consistent Skyrme-Hartree-Fock + BCS calculations.

In our previous works
\cite{Gaidarov2011,Gaidarov2012,Gaidarov2014,Antonov2017,Antonov2016} the CDFM
allowed us to make the transition from nuclear matter to finite nuclei in the
studies of the nuclear symmetry energy (NSE) for spherical \cite{Gaidarov2011}
and deformed \cite{Gaidarov2012} nuclei, as well as for Mg isotopes
\cite{Gaidarov2014} using the Brueckner energy-density functional (EDF) of
asymmetric nuclear matter \cite{Brueckner1968}. In our work \cite{Antonov2017}
we used a similar method to investigate the temperature dependence of the NSE
for isotopic chains of even-even Ni, Sn, and Pb nuclei following the local
density approximation \cite{Agrawal2014,Samaddar2007,Samaddar2008,De2012} and
using instead of the Brueckner EDF, the Skyrme EDF with SkM* and SLy4 forces. In
our work \cite{Antonov2016} the volume and surface contributions to the NSE and
their ratio were calculated within the CDFM using two EDF's, namely the
Brueckner \cite{Brueckner1968} and Skyrme (see Ref.~\cite{Wang2015}) ones.
Recently our results for the mentioned quantities have been given in
Ref.~\cite{Antonov2018}.

The structure of this article is the following. In Sec.~\ref{sec:2} we present
the common definitions of the symmetry energy and properties of nuclear matter
which characterize its density dependence around normal nuclear matter density.
A brief description of the non-relativistic Brueckner-Hartree-Fock formalism for
calculating the ground-state properties of symmetric nuclear matter and
extracting the density dependence of the nuclear symmetry energy in quadratic
approximation of the isospin polarization is given, as well. In Sec.~\ref{sec:3}
we present a brief account of the formalism of the CDFM which allows us to
relate the intrinsic quantities in nuclear matter to their corresponding ones in
finite nuclei using different phenomenological potentials. In Sec.~\ref{sec:4}
we present and discuss the results of our calculations of the already mentioned
quantities of spherical nuclei for three different isotopic chains of even-even
Ni ($A$=74--84), Sn ($A$=124--152), and Pb ($A$=206--214) nuclei within the CDFM
formalism in the framework of Skyrme HF + BCS obtained from two different
nuclear potentials for nuclear matter interactions: on one hand the modern
realistic Bonn B and Bonn CD potentials \cite{Machleidt_ANP1989,Machleidt2001}
and on the other the time-honored Brueckner's density dependent potential
\cite{Brueckner1968,Brueckner1969}. The effect of the Bonn B and Bonn CD
potentials plus the microscopic three-body forces (TBF) on the symmetry energy
and related quantities on the example of Ni isotopes is also estimated and
discussed. The conclusions of the present work are given in Sec.~\ref{sec:5}.

\section{The key EOS parameters in nuclear matter and in finite nuclei 
\label{sec:2}}
\subsection{Semi-empirical Bethe-Weizs\"{a}cker formula for nuclear matter}

The semi-empirical Bethe-Weizs\"{a}cker formula
\cite{vonWeiz35,Bethe71} captures the essential dependence of the
finite nucleus ground state on isospin asymmetry (polarization).
This formula may be viewed as a Taylor series expansion in the
energy per particle for nuclear matter in terms of the isospin
asymmetry $\delta=(\rho_{n}-\rho_{p})/\rho$, in which the
density-dependent coefficient in front of the quadratic term
defines the so-called symmetry energy $S^{NM}(\rho)$
\begin{equation}
E(\rho,\delta)=E(\rho,0)+S^{NM}(\rho)\delta^2+O(\delta^4)+
\cdots ,
\label{eq:1}
\end{equation}
where $\rho=\rho_{n}+\rho_{p}$ is the baryon density with $\rho_{n}$ and
$\rho_{p}$ denoting the neutron and proton densities, respectively (see, e.g.,
\cite{Diep2003,Chen2011}). Odd powers of $\delta$ are forbidden by the isospin
symmetry and the terms proportional to $\delta^{4}$ and higher orders are found
to be negligible.

Near the saturation density $\rho_{0}$ the energy of isospin-symmetric matter
$E(\rho,0)$ and the symmetry energy, $S^{NM}(\rho)$, can be expanded as
\begin{equation}
E(\rho,0)=E_{0}+\frac{K^{NM}}{18\rho_{0}^{2}}(\rho-\rho_{0})^{2}+
\cdots
\label{eq:2}
\end{equation}
and
\begin{eqnarray}
S^{NM}(\rho)&=&\frac{1}{2}\left.
\frac{\partial^{2}E(\rho,\delta)}{\partial\delta^{2}} \right
|_{\delta=0} = a_{4}+\frac{p^{NM}_{0}}{\rho_{0}^{2}}(\rho-\rho_0)\nonumber \\
&&+ \frac{\Delta
K^{NM}}{18\rho_{0}^{2}}(\rho-\rho_{0})^{2}+ \cdots .
\label{eq:3}
\end{eqnarray}
The parameter $a_{4}$ is the symmetry energy at equilibrium ($\rho=\rho_{0}$).
The pressure $p_{0}^{NM}$
\begin{equation}
p_{0}^{NM}=\rho_{0}^{2}\left.
\frac{\partial{S^{NM}}}{\partial{\rho}} \right |_{\rho=\rho_{0}}
\label{eq:4}
\end{equation}
and the curvature $\Delta K^{NM}$
\begin{equation}
\Delta K^{NM}=9\rho_{0}^{2}\left.
\frac{\partial^{2}S^{NM}}{\partial\rho^{2}} \right
|_{\rho=\rho_{0}}
\label{eq:5}
\end{equation}
of the nuclear symmetry energy at $\rho_{0}$ govern its density dependence and
thus provide important information on the properties of the nuclear symmetry
energy at both high and low densities.

The Bethe-Weizs\"{a}cker semi-empirical mass formula for the
nuclear ground-state energy per nucleon describes both properties
of (infinite) nuclear matter as well as finite nuclei
\cite{Steiner2005,Myers1969}:
\begin{eqnarray}
E(A,Z)&=&-B+E_SA^{-1/3} + S(A)\frac{(N-Z)^2}{A^2}+ E_C \frac{Z^2}{A^{4/3}}
\nonumber\\
&& +E_{dif} \frac{Z^2}{A^2}+E_{ex}\frac{Z^{4/3}}{A^{4/3}}+a\Delta A^{-3/2},
\label{eq:6}
\end{eqnarray}
where
\begin{equation}
S(A)=\frac{S^V(A)}{1 + \displaystyle \frac{S^S(A)}{S^V(A)}A^{-1/3}}=
\frac{S^V(A)}{1+ \displaystyle\frac{A^{-1/3}}{\kappa(A)}} \label{eq:7}
\end{equation}
with
\begin{equation}
\kappa (A) \equiv \frac{S^V(A)}{S^S(A)}.
\label{eq:8}
\end{equation}
In Eq.~(\ref{eq:6}) $B \simeq 16$ MeV is the binding energy per particle of bulk
symmetric matter at saturation. $ E_S $, $ E_C $, $ E_{dif} $, and $ E_{ex} $
are coefficients that correspond to the surface energy of symmetric matter, the
Coulomb energy of a uniformly charged sphere, the diffuseness correction, and
the exchange correction to the Coulomb energy, while the last term gives the
pairing corrections ($\Delta$ is a constant and $a=+1$ for odd-odd nuclei, 0 for
odd-even, and -1 for even-even nuclei). $S^V$ is the volume symmetry energy
parameter and $S^S$ is the modified surface symmetry energy in the droplet model
(see Ref.~\cite{Steiner2005}, where it is defined by $S^{S*}$).

In our previous work \cite{Antonov2017} we have studied the temperature
dependence of the nuclear symmetry energy $S(A,T)$, including its volume and
surface \cite{Antonov2018} contributions as derived from Brueckner
\cite{Brueckner1968,Brueckner1969} and Skyrme \cite{Wang2015} energy density
functionals. In the present work we extract the symmetry energy coefficients in
Ni, Sn, and Pb from many-body Brueckner-Hartree-Fock ground-state calculations
of nuclear matter with realistic potentials (Bonn B and Bonn CD) at zero
temperature ($T=0$ MeV) by applying the CDFM using the finite nuclei densities
obtained in self-consistent Hartree-Fock + BCS calculations with Skyrme
effective interactions (SLy4, Sk3, and SGII).

Here we would like to comment on the following point. In the present work (and
also in our previous paper \cite{Antonov2018}) we use Eq.~(\ref{eq:7}) as a
relation between the symmetry energy $S(A)$ and its volume $S^{V}(A)$ and
surface $S^{S}(A)$ components. As mentioned above, this relation is given in the
droplet model, e.g., in Ref.~\cite{Steiner2005}. However, Eq.~(\ref{eq:7}) is
different from the relation in another approach used in e.g.,
Refs.~\cite{Dan2003,Danielewicz,Dan2006,Dan2004}, and also in our work
\cite{Antonov2016}. In the latter works the relation between $S(A)$ and its
components $S^{V}(A)$ and $S^{S}(A)$ is $S^V(A)/\{1
+[S^{V}(A)/S^{S}(A)]A^{-1/3}\}$, which contains a ratio in the second term in
the denominator $[S^{V}(A)/S^{S}(A)]A^{-1/3}$, while in Eq.~(\ref{eq:7}) the
ratio is $[S^{S}(A)/S^{V}(A)]A^{-1/3}$. In the nuclear matter limit, when $A
\rightarrow \infty$ and $S^S/S^V \rightarrow 0$, the symmetry energy in
Eq.~(\ref{eq:7}) has the correct limit $S\rightarrow S^V$. In this limit the
ratio $[S^{V}(A)/S^{S}(A)]A^{-1/3}$ is not well determined and to get the right
nuclear matter limit one has to impose the condition that the surface
coefficient $S^S(A)$ goes to zero more slowly than $A^{-1/3}$ as $A \rightarrow
\infty$. To avoid this constraint we adopt in our present work (and also in
Ref.~\cite{Antonov2018}) the relation from the droplet model [Eq.~(\ref{eq:7})].

Also, at large $A$ Eq.~(\ref{eq:7}) can be written in the known form (see
Ref.~\cite{Bethe71}):
\begin{equation}
S(A) \simeq S^V-\frac{S^S}{A^{1/3}}, \label{eq:9}
\end{equation}
From Eqs.~(\ref{eq:7}) and (\ref{eq:8}) follow the relations of $S^V(A)$ and
$S^S(A)$ with $S(A)$:
\begin{equation}
S^V(A)=S(A)\left(1+\frac{1}{\kappa(A) A^{1/3}}\right),
\label{eq:10}
\end{equation}
\begin{equation}
S^S(A)=\frac{S(A)}{\kappa(A)}\left(1+\frac{1}{\kappa (A) A^{1/3}}\right) .
\label{eq:11}
\end{equation}

Due to the choice of Eq.~(\ref{eq:7}), Eqs.~(\ref{eq:10}) and (\ref{eq:11}) are
different from Eqs.~(\ref{eq:26}) and (\ref{eq:27}) in Ref.~\cite{Antonov2016}.
This leads to different results for $S^{V}(A)$ and $S^{S}(A)$ obtained in
\cite{Antonov2016} and those of the present work which will be shown in
Sec.~\ref{sec:4}. We emphasize, however, that the results for the symmetry
energy $S(A)$ in this work, as well as in Refs.~\cite{Antonov2016} and
\cite{Gaidarov2011}, should be the same because they are obtained from identical
equations, namely, Eq.~(\ref{eq:27}) in Sec.~\ref{sec:3} of the present work,
Eq.~(20) in \cite{Antonov2016}, and Eq.~(28) in Ref.~\cite{Gaidarov2011}. As can
be expected, the differences concern the values of $S^{V}(A)$ and $S^{S}(A)$.

Furthermore, it has been argued (Refs.~\cite{Dan2006,Diep2007}) that the ratio
of the volume to the surface energy coefficients is given by the following
integral of the symmetry energy function of density $S(\rho)$:
\begin {equation}
\kappa (A)=\frac{S^V(A)}{S^S(A)}=\frac{3}{r_{0}}\int dr \frac{\rho(r)}{\rho_{0}}
\left \{\frac{S^{NM}(\rho_{0})}{S^{NM}[\rho(r)]}-1\right \} \label{eq:12}
\end{equation}
in the local density approximation to the symmetry energy. In Eq.~(\ref{eq:12})
$\rho(r)$ is the half-infinite nuclear matter density, $\rho_{0}$ is the nuclear
matter equilibrium density, and $r_{0}$ is the radius of the nuclear volume per
nucleon. The latter two quantities are related by
\begin {equation}
\frac{4\pi r_{0}^{3}}{3}=\frac{1}{\rho_{0}}.
\label{eq:13}
\end{equation}

In Section~\ref{sec:4} are presented our results for the volume
and surface components of the nuclear symmetry energy and their
ratio obtained in calculations based on relationships from this
section and within the CDFM (see Section~\ref{sec:3}).

\subsection{Ground-state properties of dense nuclear matter calculated with Bonn
B and Bonn CD nuclear potentials}

In this subsection we give a short overview of the formalism employed in
obtaining the density dependence of the symmetry energy of nuclear matter.
Details can be seen in the original works of Engvik {\it et
al.}~\cite{Engvik1997}, and also of Haftel and Tabakin
\cite{Haftel-Tabakin1970}, Erkelenz {\it et al.}
\cite{Erkelenz-Holinde-Bleuler1971}, Machleidt
\cite{Machleidt_ANP1989,Machleidt_CompNuc2}, and Song {\it et al.}
\cite{Song1991,Song1992}. Excellent pedagogical reviews can be found in M. Baldo
\cite{Baldo1999} and the lectures of Hjorth-Jensen \cite{Hjorth-Jensen2000} at
the Twelfth Summer School in Nuclear Physics at the University of California,
Santa Cruz in 2000.

The symmetry energy in nuclear matter ${S}^{NM} (\rho)$ as a function of density
(neglecting higher than quadratic terms in isospin asymmetry) can be
approximated with
\begin{equation}
{S}^{NM} (\rho) = {\cal E} (\rho,\delta=1) - {\cal E} (\rho,\delta=0 ),
\label{eq:14}
\end{equation}
where ${\cal E} (\rho,\delta=1) $ is the ground state energy per nucleon of
completely isospin polarized (i.e. neutron matter) at density $ \rho$ and ${\cal
E} (\rho,\delta=0)$ is the ground-state energy per nucleon of isospin
unpolarized (i.e. isospin symmetric) matter. From Eqs.~(\ref{eq:4}) and
(\ref{eq:14}) it follows the relationship for the pressure $p_0^{NM}(\rho)$.

The calculation of the ground-state energies per nucleon at different density
and polarization is carried out using the reaction matrix $G$, which is a
solution of the Bethe-Goldstone equation for various isospin polarization
fractions $\delta$:
\begin{equation}
G(\omega ,\delta)=V+V\frac{Q(\delta)}{\omega - H_0}G(\omega ,\delta),
\label{eq:15}
\end{equation}
where $\omega$ is the unperturbed energy of the interacting nucleons, $V$ is the
free $NN$ potential, $H_0$ is the unperturbed energy of the intermediate
scattering states, and $Q(\delta)$ is the Pauli operator preventing scattering
into occupied states. Only ladder diagrams with intermediate two-particle states
are included in Eq.~(\ref{eq:15}).

\begin{figure}[htb]
\centering
\includegraphics[width=0.9\columnwidth]{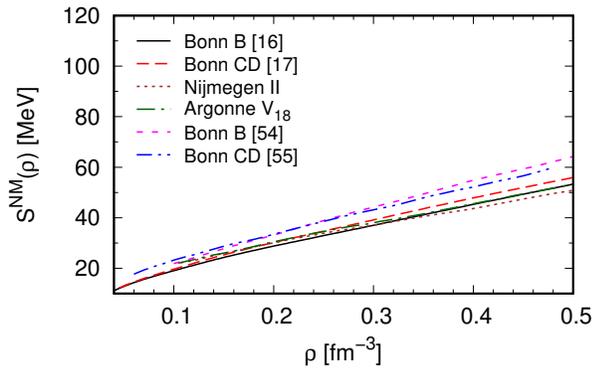}
\caption[]{Density dependence of the symmetry energy calculated with Bonn B and
Bonn CD potentials from \cite{Machleidt_ANP1989,Machleidt2001}, as well as
from \cite{Wang2014,Soma2008} and with Nijmegen II and Argonne $V_{18}$ potentials
from \cite{Engvik1997}. We note that all the curves correspond to two-body forces.
\label{fig.1}}
\end{figure}

\begin{figure}[htb]
\centering
\includegraphics[width=0.9\columnwidth]{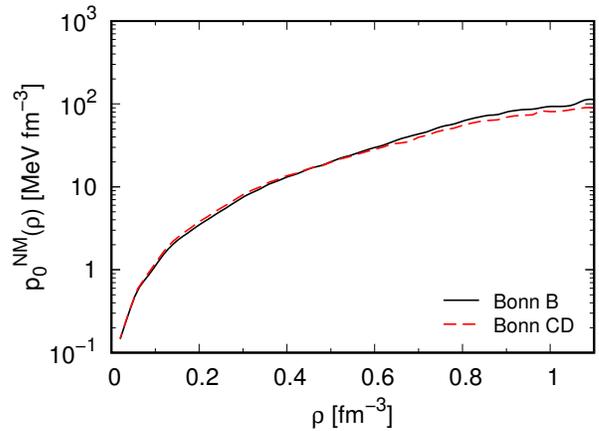}
\caption[]{Density dependence of the pressure in nuclear matter calculated with
Bonn B and Bonn CD potentials from \cite{Machleidt_ANP1989,Machleidt2001}.
\label{fig.2}}
\end{figure}

Having obtained a self-consistent solution of Bethe-Goldstone equation, one can
construct the bound state energies per nucleon at different density and isospin
asymmetry (polarization) $\delta$
\begin{equation}
{\cal E}(\rho,\delta)={\cal T}+{\cal U}(\rho,\delta)
\label{eq:16}
\end{equation}
with the kinetic energy
\begin{equation}
{\cal T} = \frac{3}{10m k_F^3} (k_{Fp}^5 + k_{Fn}^5), \label{eq:17}
\end{equation}
where $m$ is the effective nucleon mass and $k_F$ is the total Fermi momentum.
The Fermi momenta $k_{Fp}$ and $k_{Fn}$ for protons and neutrons, respectively,
are related to the total nuclear density $\rho$ and isospin asymmetry
(polarization) $\delta$ by
\begin{eqnarray}
\rho & = & \frac{2}{3\pi^2} k_F^3 = \frac{(1+\delta)}{2}\rho + \frac{(1-\delta)}{2}\rho \nonumber \\
& = & \frac{1}{3\pi^2} k_{Fn}^3 + \frac{1}{3\pi^2} k_{Fp}^3. \label{eq:18}
\end{eqnarray}
The contribution of the potential energy ${\cal U}$ to the total energy per
particle can be written in the form:
\begin{eqnarray}
{\cal U}(\rho,\delta)& = &\frac{1}{2\rho}\frac{1}{(2\pi)^6}\sum _{a,b=(pn)}
\int_0^{k_{Fa}} d^3k_a \int_0^{k_{Fb}} d^3k_b \nonumber\\
& \times & \langle k_ak_b\vert G(\omega =
\epsilon _a(k_a) + \epsilon _b(k_b))\vert k_ak_b\rangle,
\label{eq:19}
\end{eqnarray}
where $\epsilon_a$ and $\epsilon_b$ are the nucleon single-particle energies.
Using this approach we have carried out first non-relativistic
Brueckner-Hartree-Fock calculations of the symmetry energy using Bonn B and Bonn
CD potentials from \cite{Machleidt_ANP1989,Machleidt2001}. Our results are
summarized in Fig.~\ref{fig.1}. They are in line with previous studies which
confirm the nearly linear dependence of the symmetry energy in nuclear matter
with density \cite{Engvik1997,CHLee1998}. In the same figure are presented also
the results when two-body Bonn B potential from Ref.~\cite{Wang2014} and
two-body Bonn CD potential from Ref.~\cite{Soma2008}, as well as the results for
$S^{NM}(\rho)$ obtained by the Nijmegen II and Argonne $V_{18}$ \cite{Akmal1997}
potentials taken from Ref.~\cite{Engvik1997} are used. We note that in
Fig.~\ref{fig.1} the TBF are not included. Here we would like to emphasize that
when comparing the results in Fig.~\ref{fig.1} one has to keep in mind their
dependence on the choices of the auxiliary potentials and of the methods of
calculations. In the paper of Wang \textit{et al.} \cite{Wang2014} BHF
calculations are performed (with Bonn B two-body potential and microscopic TBF)
using gap and continuous auxiliary potentials that lead to different results,
namely, the symmetry energy under the gap choice is smaller than that under the
continuous choice. In Fig.~\ref{fig.1} the result of \cite{Wang2014} with a gap
auxiliary potential is compared with our result using also gap auxiliary
potential and Bonn B potential from \cite{Machleidt_ANP1989}. We should note the
existing small difference between our result and that from Ref.~\cite{Wang2014}.
The reason for this is mainly the difference between the estimations made in our
work and those in Ref.~\cite{Wang2014} of the contribution of the $^1P_1$
partial wave to the energy as a function of the density. For instance, for the
density $\rho=0.17$ fm$^{-3}$ our estimation is close to that in Table 9.2 of
Ref. [16] but it is about 3 MeV larger than that in Table 1 of
Ref.~\cite{Wang2014}. We should add, however, that the noted difference is
within the uncertainty in the experimental value of the symmetry energy (e.g.
$S= 32 \pm 6$ MeV given in \cite{Soma2008}). The calculations of Soma and Bozek
\cite{Soma2008} are performed not within the BHF method but using the
self-consistent in-medium T-matrix approach implemented with Bonn CD and
Nijmegen potentials plus the three-nucleon Urbana interaction. It is found in
Ref.~\cite{Soma2008} that the results of their approach for the symmetry energy
in the case of the Bonn CD potential is $S=30$ - $32$ MeV. In Fig.~\ref{fig.2}
are given the results of our calculations for the pressure in nuclear matter
based on Eqs.~(\ref{eq:4}) and (\ref{eq:14}) for both Bonn potentials.

\begin{figure}[htb]
\centering
\includegraphics[width=0.9\columnwidth]{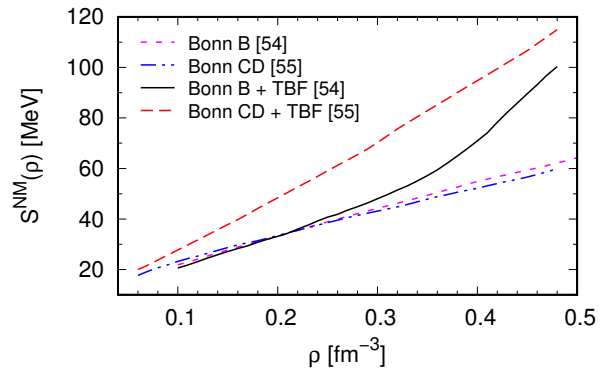}
\caption[]{Density dependence of the symmetry energy calculated with Bonn B
\cite{Wang2014} and Bonn CD \cite{Soma2008} potentials with two- and three-body
forces (TBF).
\label{fig.3}}
\end{figure}

In the end of this section we would like to comment on the role of the
three-body forces. It is well known that the standard BHF formalism involving
only two-body forces does not reproduce correctly the empirical saturation point
of nuclear matter \cite{Coester70}. The studies of the TBF effects on the
properties of symmetric NM and pure neutron matter in
Refs.~\cite{Baldo2008a,Wang2014,Soma2008} were performed for Argonne $V_{18}$,
CD Bonn, and Bonn B two-body potentials, which are among the few most accurate
NN interactions. The latter two potentials are object of the present work and,
in principle, of primary interest to investigate the dependence of the nuclear
EOS on the two- and three-body forces and their impact on the characteristics of
NM for finite nuclei. Indeed, the saturation density and the energy per particle
of nuclear matter can be improved by including the TBF to (0.17 fm$^{-3}$ and
-15.9 MeV) \cite{Wang2014} and to (0.185 fm$^{-3}$ and -15.5 MeV)
\cite{Baldo2008a} for the Bonn B and Bonn CD potentials, respectively. For
comparison, the corresponding saturation points in the case of Brueckner EDF are
(0.204 fm$^{-3}$ and -16.57 MeV). In Fig.~\ref{fig.3} are compared the results
for the symmetry energy as a function of the density from the calculations
within the BHF method for Bonn B potential \cite{Wang2014} (with a gap choice)
and within the in-medium T-matrix approach for Bonn CD potential \cite{Soma2008}
with and without TBF. In general, as it can be seen from Fig.~\ref{fig.3}, the
TBF play an important role in determining the high-density behavior of the
symmetry energy and its effect leads to a strong stiffening of the symmetry
energy at high densities. A confirmation of this fact can be also found in
Ref.~\cite{Gandolfi2012}, where the TBF play a key role in determining the
coefficient $b$ [see Eq.~(3) and Table I of Ref.~\cite{Gandolfi2012}], which is
responsible for the high-density behavior of the EOS and whose values vary
significantly when two-body or three-body forces are accounted for in quantum
Monte Carlo calculations of the neutron star mass-radius relationship.

\section{Nuclear EOS parameters of finite nuclei in the CDFM
\label{sec:3}}

In what follows we calculate the key EOS parameters in finite nuclei, i.e. the
pressure $p_{0}$ at saturation density $\rho_0$ and the nuclear symmetry energy
and its surface and volume components (see
Refs.~\cite{Antonov80,AHP,Antonov2016}) using the CDFM. The latter is a natural
extension of the Fermi-gas model and is based on the $\delta$-function
approximation of the generator coordinate method \cite{Grif57}. The model
includes nucleon-nucleon correlations of collective type. In general, it allows
us to make the transition from quantities in nuclear matter to the corresponding
ones in finite nuclei. In the present work it is applied to our studies of the
symmetry energy and its components. In the CDFM the one-body density matrix
$\rho(\mathbf{r},\mathbf{r}^{\prime})$ is a coherent superposition of the
one-body density matrices $\rho^{NM}_{x}({\bf r},{\bf r^{\prime}})$ for
spherical ``pieces'' of nuclear matter (``fluctons'') with densities
$\rho_{x}({\bf r})=\rho_{0}(x)\Theta(x-|{\bf r}|)$ and $\rho_{0}(x)=3A/4\pi
x^{3}$. It has the form:
\begin{equation}
\rho({\bf r},{\bf r^{\prime}})=\int_{0}^{\infty}dx |{\cal
F}(x)|^{2} \rho^{NM}_{x}({\bf r},{\bf r^{\prime}})
\label{eq:20}
\end{equation}
with
\begin{eqnarray}
\rho^{NM}_{x}({\bf r},{\bf r^{\prime}})&=&3\rho_{0}(x) \frac{j_{1}(k_{F}(x)
|{\bf r}-{\bf r^{\prime}}|)}{(k_{F}(x)|{\bf r}-{\bf r^{\prime}}|)}\nonumber \\
 & \times & \Theta \left (x-\frac{|{\bf r}+{\bf r^{\prime}}|}{2}\right ).
\label{eq:21}
\end{eqnarray}
In (\ref{eq:21}) $j_{1}$ is the first-order spherical Bessel function and
\begin{equation}
k_{F}(x)=\left(\frac{3\pi^{2}}{2}\rho_{0}(x)\right )^{1/3}\equiv
\frac{\beta}{x}
\label{eq:22}
\end{equation}
with
\begin{equation}
\beta=\left(\frac{9\pi A}{8}\right )^{1/3}\simeq 1.52A^{1/3}
\label{eq:23}
\end{equation}
is the Fermi momentum of the nucleons in the flucton with a radius $x$. The
density distribution in the CDFM has the form:
\begin{equation}
\rho({\bf r})=\int_{0}^{\infty}dx|{\cal
F}(x)|^{2}\rho_{0}(x)\Theta(x-|{\bf r}|).
\label{eq:24}
\end{equation}
It follows from (\ref{eq:24}) that in the case of monotonically decreasing local
density ($d\rho/dr\leq 0$) the weight function $|{\cal F}(x)|^{2}$ can be
obtained from a known density (theoretically or experimentally obtained):
\begin{equation}
|{\cal F}(x)|^{2}=-\frac{1}{\rho_{0}(x)} \left. \frac{d\rho(r)}{dr}
\right |_{r=x}
\label{eq:25}
\end{equation}
with normalization
\begin{equation}
\int_{0}^{\infty}dx |{\cal F}(x)|^{2}=1.
\label{eq:26}
\end{equation}

We have shown in our previous works \cite{Gaidarov2011,Gaidarov2012,Antonov2016}
by applying the CDFM that both nuclear symmetry energy and pressure in finite
nuclei can be obtained from infinite nuclear matter at temperature $T=0$ MeV by
weighting it with $|{\cal F}(x)|^{2}$ :
\begin{equation}
S(A)=\int_{0}^{\infty}dx|{\cal F}(x)|^{2}S^{NM}[\rho(x)],
\label{eq:27}
\end{equation}
\begin{equation}
p_{0}(A)=\int_{0}^{\infty}dx|{\cal F}(x)|^{2}p_{0}^{NM}[\rho(x)].
\label{eq:28}
\end{equation}

Self-consistency requires that when this procedure is applied to quantities
referring to (infinite) nuclear matter, the weight function reduces to Dirac
delta function. For example, when the condition for self-consistency is applied
to the density $\rho(|{\bf r}|)$ and the symmetry energy $S^{NM}[\rho(|{\bf
r}|)]$ in nuclear matter it leads from Eq.~(\ref{eq:24}) and Eq.~(\ref{eq:27})
to the trivial identities:
\begin{eqnarray}
\rho^{NM}(|{\bf r}|,x)&=&\int_{0}^{\infty}dx' \delta(x'-x)\rho_{0}(x')\Theta(x'-|{\bf r}|)\nonumber\\
&=& \rho_{0}(x)\Theta(x-|{\bf r}|).
\label{eq:27.1}
\end{eqnarray}
\begin{eqnarray}
S^{NM}[\rho^{NM}(|{\bf r}|,x)]&=&\int_{0}^{\infty}dx' \delta(x'-x)S^{NM}[\rho^{NM}(|{\bf r}|,x')]\nonumber\\
&=&S^{NM}[\rho_{0}(x)\Theta(x-|{\bf r}|)],
\label{eq:27.2}
\end{eqnarray}

In Refs.~\cite{Antonov2016,Gaidarov2011,Gaidarov2012} we used expressions for
$S^{NM}[\rho(x)]$ and $p_{0}^{NM}[\rho(x)]$ derived for nuclear matter from an
effective density-dependent Brueckner potential
\cite{Brueckner1968,Brueckner1969}:
\begin{eqnarray}
S^{NM}(x)&=&41.7\rho_{0}^{2/3}(x)+b_{4}\rho_{0}(x) \nonumber \\
&& + b_{5}\rho_{0}^{4/3}(x)+b_{6}\rho_{0}^{5/3}(x),
\label{eq:29}
\end{eqnarray}
\begin{eqnarray}
p_{0}^{NM}(x)&=&27.8\rho_{0}^{5/3}(x)+b_{4}\rho_{0}^{2}(x) \nonumber \\
&& + \frac{4}{3}b_{5}\rho_{0}^{7/3}(x)+\frac{5}{3}b_{6}\rho_{0}^{8/3}(x),
\label{eq:30}
\end{eqnarray}
where
\begin{eqnarray}
b_{4}&=&148.26, \;\;\;\; b_{5}=372.84, \;\;\;\; b_{6}=-769.57.
\label{eq:31}
\end{eqnarray}

In the present work using the CDFM we take nuclear matter values of the
considered parameters and use them to deduce the corresponding values in finite
nuclei. We apply the CDFM in the framework of a self-consistent
Skyrme-Hartree-Fock plus BCS method to calculate the volume and surface
contributions to the symmetry energy and their ratio, as well as the pressure,
in spherical nuclei of the Ni, Sn, and Pb isotopic chains. In our approach to
calculate $\kappa(A)$, i.e the ratio $S^{V}(A)/S^{S}(A)$ (see also
Ref.~\cite{Antonov2016}) we start from Danielewicz's formula Eq.~(\ref{eq:12}).
In it we make an approximation replacing the density $\rho(r)$ for the
half-infinite nuclear matter in the integrand by the density distribution of a
finite nucleus. We replace the latter by the expression in the CDFM
[Eq.~(\ref{eq:24})]. Concerning the term $S^{NM}[\rho(r)]$, when use is made of
Eq.~(\ref{eq:27.1}) and Eq.~(\ref{eq:27.2}) the formula for $\kappa(A)$ takes
the form

\begin{eqnarray}
\kappa(A) &=&
\frac{3}{r_{0}\rho_{0}}\int_{0}^{\infty}dx
|{\cal F}(x)|^{2} \rho_{0}(x) \nonumber \\ 
&&\times\int_{0}^{x}dr\left
\{\frac{S^{NM}(\rho_{0})}{S^{NM}[\rho_0(x)]}-1\right \},
\label{eq:32.0}
\end{eqnarray}
which leads to
\begin{equation}
\kappa(A)=\frac{3}{r_{0}\rho_{0}}\int_{0}^{\infty}dx
|{\cal F}(x)|^{2} x \rho_{0}(x)\left
\{\frac{S^{NM}(\rho_{0})}{S^{NM}[\rho_0(x)]}-1\right \} ,
\label{eq:32}
\end{equation}
where the right-hand side of Eq.~(\ref{eq:32}) is an one-dimensional integral
over $x$. The latter is the radius of the ``flucton'' that is perpendicular to
the nuclear surface.

\begin{figure}
\centering
\includegraphics[width=\columnwidth]{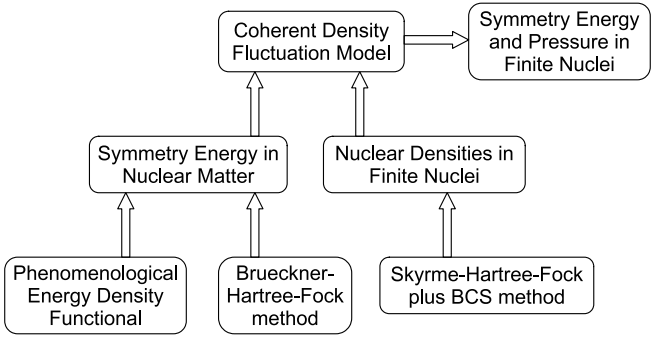}
\caption[]{Block diagram of the application of the CDFM for the extraction of
symmetry energy and pressure in finite nuclei from their counterparts in nuclear
matter.
\label{fig.4}}
\end{figure}

The weight function $|{\cal F}(x)|^{2}$ is calculated from Eq.~(\ref{eq:25})
using the finite nucleus density $\rho(r)$ obtained from self-consistent
deformed Hartree-Fock plus BCS calculations with density-dependent Skyrme
interactions. We use expressions for $S^{NM}(x)$ and $p^{NM}_0(x)$ derived using
Bonn B and Bonn CD potentials in non-relativistic Brueckner-Hartee-Fock
calculations \cite{Engvik1997} of ground-state properties of nuclear matter at
different densities. The role of the CDFM to extract the NSE and pressure in
finite nuclei from their counterparts in infinite nuclear matter is presented
schematically by a block diagram in Fig.~\ref{fig.4}.

\section{Results of calculations \label{sec:4}}

In this Section we present the obtained results for the symmetry energy and
pressure in finite nuclei extracted from nuclear matter many-body calculations
using realistic Bonn B and Bonn CD potentials. We show also the results for the
volume and surface contributions to the nuclear symmetry energy.

\begin{figure}
\centering
\includegraphics[width=0.9\columnwidth]{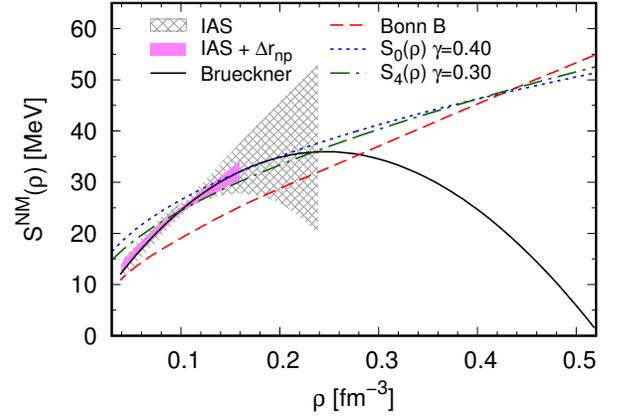}
\caption[]{
Density dependence of the symmetry energy $S^{NM}(\rho)$ in nuclear matter with
Brueckner and Bonn B potentials \cite{Machleidt_ANP1989,Machleidt2001}. The
curves for $S_0(\rho)$ ($\gamma = 0.40$) [Eq.~(\ref{eq:33})] and $S_4(\rho)$
($\gamma =0.30$) [Eq.~(\ref{eq:34})] are also given. The constraints on the
symmetry energy taken from Fig.~4 of Ref.~\cite{Logoteta2019} are presented by
two bands (see the text).\label{fig.5}}
\end{figure}

In Fig.~\ref{fig.5} we plot the symmetry energy in nuclear matter as a function
of density for both Bonn B and Brueckner effective potential (from
\cite{Machleidt_ANP1989,Machleidt2001}) in a range of nuclear densities where
the expression of symmetry energy derived from the effective Brueckner potential
is non-negative. On the same figure we present for a comparison two other
density dependencies $S^{NM}(\rho)$ used in our work \cite{Antonov2018}. These
are the power parametrization (see Refs.~\cite{Dan2004,Dan2006,Diep2007})
\begin{equation}
S^{NM}(\rho)= S^V \left( \frac{\rho}{\rho_0}\right)^{\gamma}
\label{eq:33}
\end{equation}
[noted in the figure by $S_0(\rho)$] for the case when $\gamma = 0.40$ and also
(Refs.~\cite{Tsang2009,Dong2012}):
\begin{equation}
S^{NM}(\rho)=12.5\left(\frac{\rho}{\rho_{0}}\right)^{2/3}+17.6\left(\frac{\rho}{\rho_{0}}\right)^{\gamma}
\label{eq:34}
\end{equation}
[noted in the figure by $S_4(\rho)$] for the case when $\gamma =0.30$. The data
for the symmetry energy obtained in Ref.~\cite{Dan2014} from nuclear isobaric
analog states (IAS) and from neutron skin thickness $\Delta r_{np}$ of heavy
nuclei (IAS+$\Delta r_{np}$) \cite{RocaMaza2013,Zhang2013} are also presented in
Fig.~\ref{fig.5} by grey hatched and magenta bands, respectively. The two bands are
taken from Ref.~\cite{Logoteta2019}, where the role of the three-body forces
have been studied. As can be seen from Fig.~\ref{fig.5}, the symmetry energy
calculated with Bonn B potential is slightly smaller for $\rho \leq 0.25$
fm$^{-3}$ than the ones obtained from the Brueckner EDF and using the power
parametrization [Eqs.~(\ref{eq:33}) and (\ref{eq:34})]. Nevertheless, the
behavior of all symmetry energy curves is similar and in accordance with the
empirical data \cite{Dan2014,RocaMaza2013,Zhang2013} around the normal NM
density.

\begin{figure}
\centering
\includegraphics[width=0.9\columnwidth]{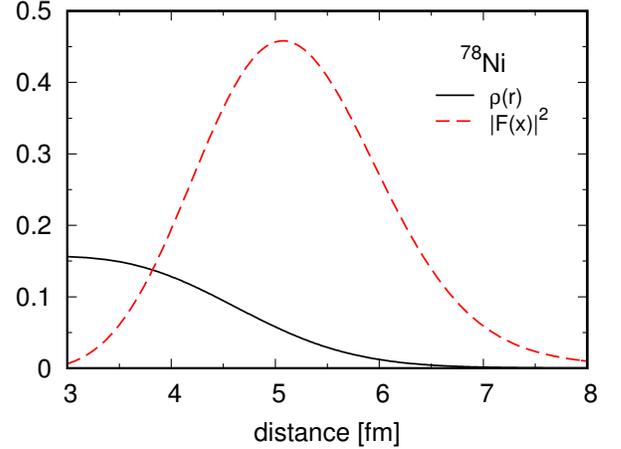}
\caption[]{The density $\rho (r)$ (in fm$^{-3}$) of $^{78}$Ni calculated in the
Skyrme HF + BCS method with SLy4 force (normalized to $A=78$) and the weight
function $|{\cal F}(x)|^{2}$ (in fm$^{-1}$) normalized to unity
[Eq.~(\ref{eq:26})]. \label{fig.6}}
\end{figure}

In Fig.~\ref{fig.6} we overlay the surface part of the density distribution of
$^{78}$Ni and the corresponding CDFM weight function $|{\cal F}(x)|^{2}$ as a
function of $x$. The density is obtained in a self-consistent Hartree-Fock + BCS
calculations with SLy4 interaction. The function $|{\cal F}(x)|^{2}$ which is
used in Eqs.~(\ref{eq:27}), (\ref{eq:28}), and (\ref{eq:32}) has the form of a
bell with a maximum around $x=R_{1/2}$ at which the value of the density
$\rho(x=R_{1/2})$ is around half of the value of the central density equal to
$\rho_0$ [$\rho(R_{1/2})/\rho_0 =0.5$]. Namely in this region around
$\rho=0.5\rho_0$ the values of different $S^{NM}(\rho)$ play the main role in
the calculations. As it is known, the central density of the nucleus has values
around $\rho_0 \approx 0.10$ - $0.16$ fm$^{-3}$. Consequently, the maximum of
the weight function $|{\cal F}(x)|^{2}$ is around $\rho(R_{1/2}) \approx 0.05$ -
$0.08$ fm$^{-3}$. In the case of $^{78}$Ni (Fig.~\ref{fig.6}) the maximum of
$|{\cal F}(x)|^{2}$ is at $\rho=0.05$ fm$^{-3}$ and, within its width range, the
density $\rho$ is between 0.12 fm$^{-3}$ and 0.01 fm$^{-3}$. Therefore, for the
finite nucleus calculations the relevant values of $S^{NM}$ are typically those
in the region around $\rho \approx 0.01$ - $0.12$ fm$^{-3}$ in Figs.~\ref{fig.1}
and \ref{fig.5}.

We apply the CDFM to derive the values of the symmetry energy and pressure in
three isotopic chains of nuclei (Ni, Sn, and Pb) from our many-body symmetry
energy calculations in nuclear matter with Bonn B and Bonn CD potentials 
from \cite{Machleidt_ANP1989,Machleidt2001} with input nuclear densities
obtained from Hartree-Fock + BCS self-consistent calculations with different
Skyrme paramatrizations. We use effective Skyrme forces SLy4 \cite{sly4}, Sk3
\cite{sk3}, and SGII \cite{sg2}, namely because they are among the most widely
used Skyrme forces and have been already used in our previous paper
\cite{Sarriguren2007}.

Using Eqs.~(\ref{eq:27}) and (\ref{eq:28}) we calculate the nuclear symmetry
energy and the pressure, correspondingly, as well as by means of
Eqs.~(\ref{eq:32}), (\ref{eq:10}), and (\ref{eq:11}) the volume and surface
contributions to the symmetry energy in finite nuclei.

\begin{figure*}[ht]
\centering
\includegraphics[width=0.8\textwidth]{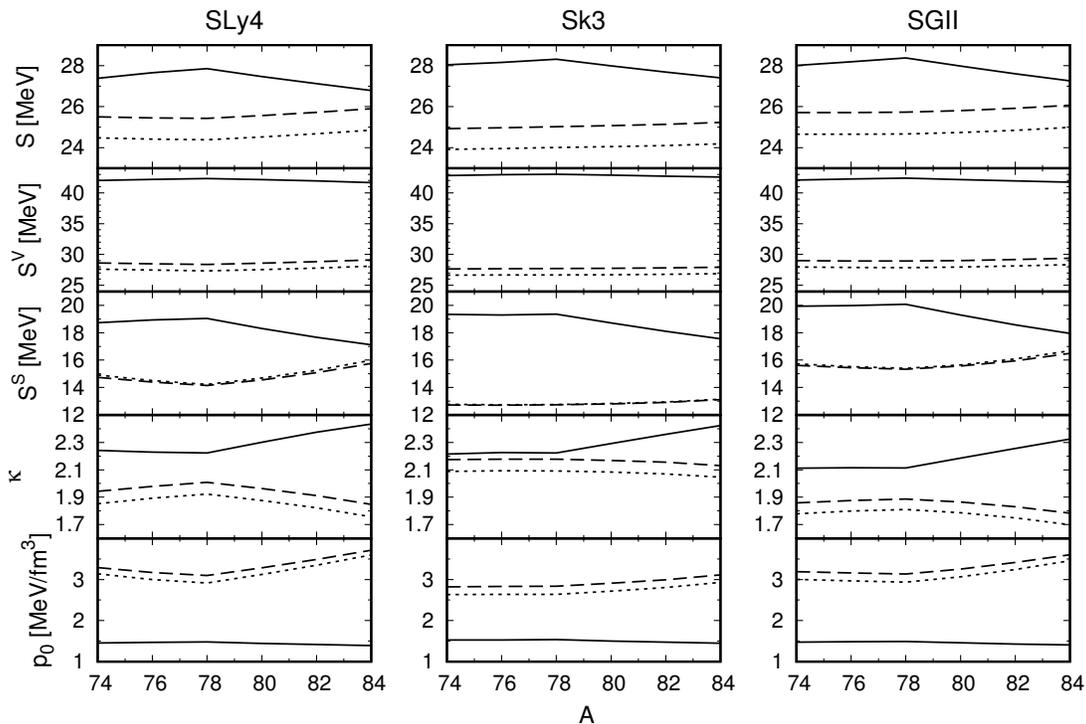}
\caption[]{The symmetry energy $S$, its volume $S^V$ and surface $S^S$
components, their ratio $\kappa$ and the pressure $p_0$ for Ni isotopes
obtained using Brueckner EDF (solid line) and BHF method with Bonn CD (dashed
line) and Bonn B (dotted line) potentials from
\cite{Machleidt_ANP1989,Machleidt2001}. The weight function $|{\cal
F}(x)|^{2}$ used in the calculations is obtained by means of the densities
derived within a self-consistent Skyrme-Hartree-Fock plus BCS method with SLy4
(left panel), Sk3 (middle panel) and SGII (right panel) Skyrme interactions.
\label{fig.7}}
\end{figure*}

\begin{figure*}
\centering
\includegraphics[width=0.8\textwidth]{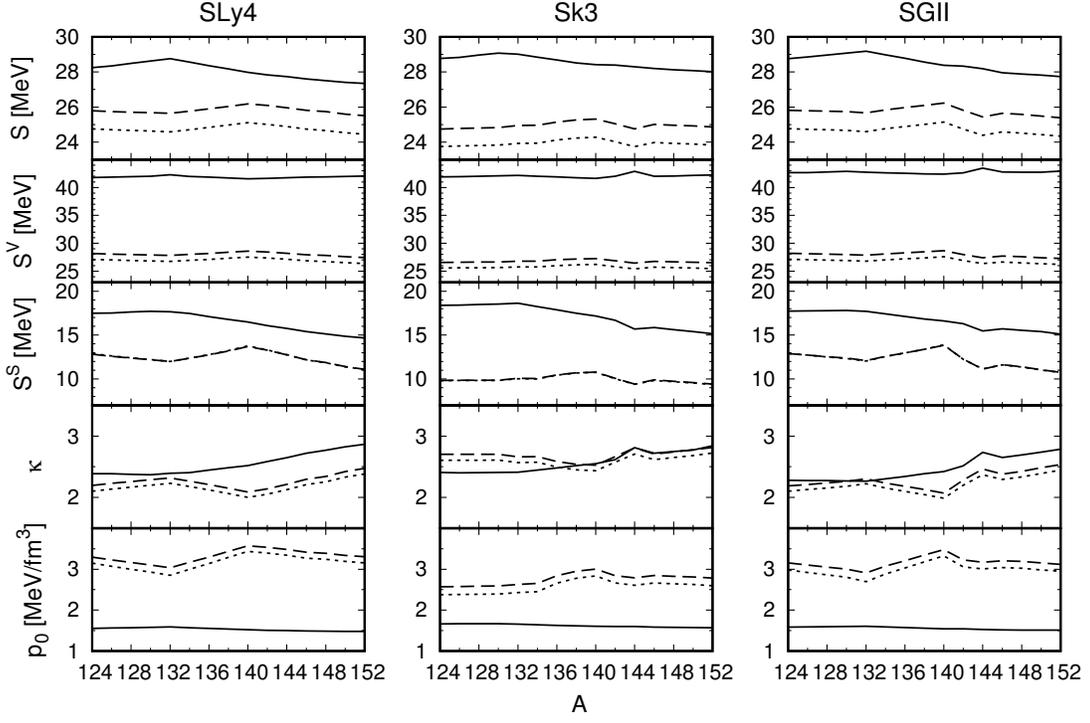}
\caption[]{Same as Fig.~\ref{fig.7}, but for Sn isotopes. \label{fig.8}}
\end{figure*}

\begin{figure*}
\centering
\includegraphics[width=0.8\textwidth]{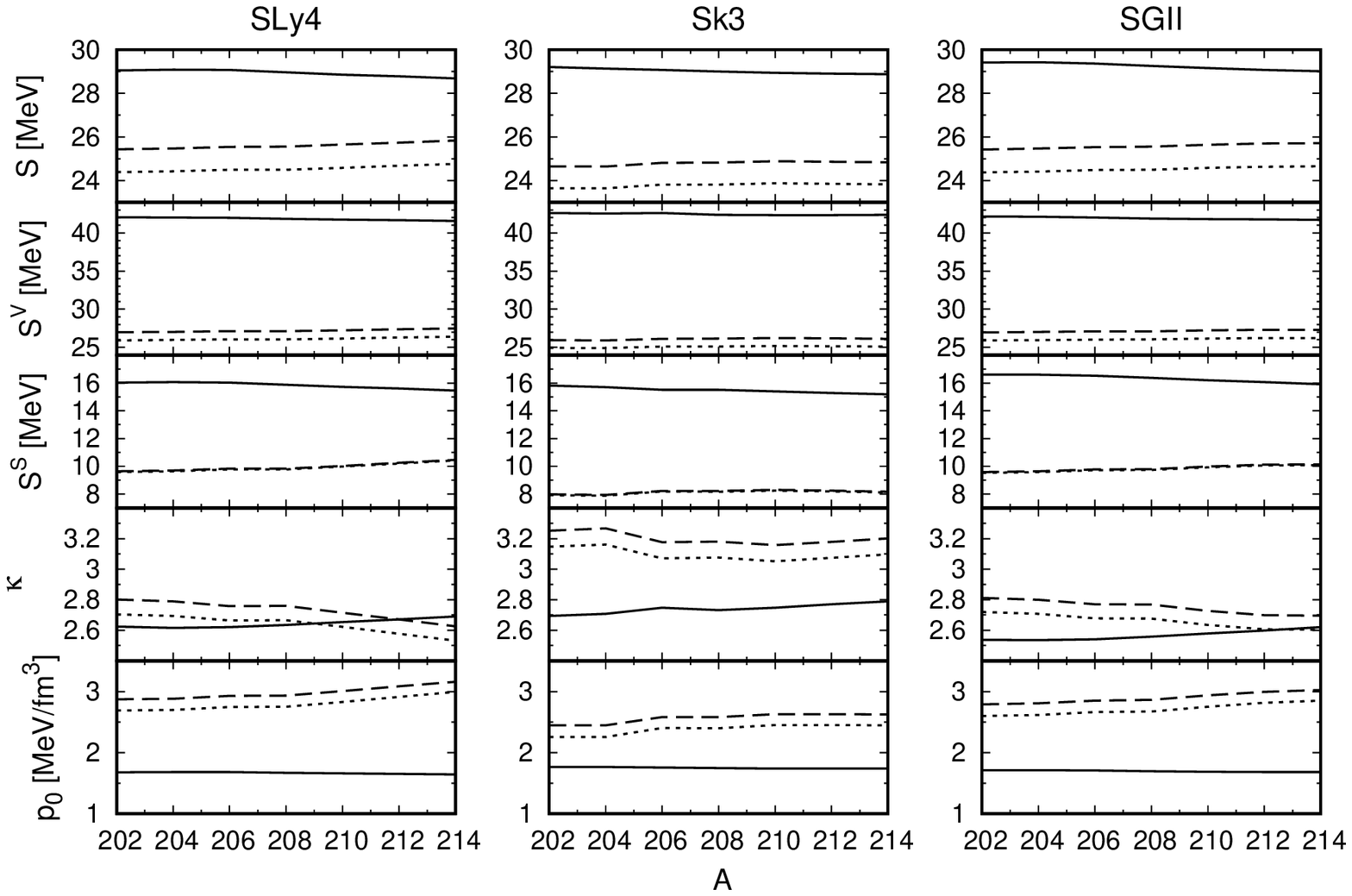}
\caption[]{Same as Fig.~\ref{fig.7}, but for Pb isotopes.
\label{fig.9}}
\end{figure*}

We next comment on the similarities and differences between our results for the
symmetry energy (along with its volume and surface components) and pressure in
the isotopes of Ni, Sn, and Pb nuclei. As can be seen in Fig.~\ref{fig.4} we use
the CDFM that links the microscopic description of finite nuclei obtained from
Skyrme-Hartree-Fock plus BCS method with the symmetry of nuclear matter. In this
work the latter is obtained in two different ways, namely by self-consistent
many-body Brueckner-Hartree-Fock calculations with realistic Bonn B and Bonn CD
potentials or by effective Brueckner density functional. In this way the CDFM
used in our hybrid approach utilizes the strong sides of both the {\it ab
initio} many-body Brueckner-Hartree-Fock calculations using Bonn B and Bonn CD
potentials (which give reasonable results for the binding energy in nuclear
matter at saturation densities) and the Skyrme-Hartree-Fock + BCS method which
gives realistic results for the nuclear density profiles in finite nuclei with
appropriate choice of the parameters of the density-dependent Skyrme contact
interaction (SLy4, Sk3, and SGII).

In the case of the Ni isotopic chain (Fig.~\ref{fig.7}) for the choice of the
Brueckner EDF the calculated symmetry energy (using the CDFM with the SLy4
force) is in the range 26.8-27.8 MeV. The volume symmetry energy $S^V$ is within
41.7-42.3 MeV and the surface symmetry energy $S^S$ in the range 17-19 MeV. The
pressure is in the range 1.39-1.48 MeV/fm$^3$ and the ratio $\kappa$ within
2.22-2.44. By inspecting the graphs (Fig.~\ref{fig.7}) of the values of these
nuclear state parameters in Ni isotopes one notices that at $A=78$ the symmetry
energy $S(A)$, as well as its volume and surface components, $S^V$ and $S^S$ and
the pressure $p_0$, reach maximum values while $\kappa$ has a minimum
($\kappa$=2.22).

For the choice of the Bonn CD one-boson exchange potential \cite{Machleidt2001}
the calculated symmetry energy (using the CDFM with the SLy4 force) is in the
range of 25.4-25.9 MeV with volume symmetry energy $S^V$ within 28.4-29.1 MeV
and surface symmetry energy $S^S$ in the range 14.1-15.6 MeV. The pressure is in
the range 3.1-3.7 MeV/fm$^3$ and the ratio $\kappa$ within 1.8-2.0. As can be
seen in Fig.~\ref{fig.5}, at $A=78$ the symmetry energy $S(A)$, as well as its
volume and surface components, $S^V$ and $S^S$ and pressure $p_0$, reach minimum
values while $\kappa$ has a weakly expressed maximum ($\kappa$=2.0).

In the case of the Bonn B one-boson exchange potential \cite{Machleidt_ANP1989}
the result for the symmetry energy (using the CDFM with the SLy4 force) is in
the range of 24.4-24.8 MeV, with volume symmetry energy $S^V$ within 27.6 and
28.1 MeV and surface symmetry energy $S^S$ within 14.9 and 16.0 MeV. The
pressure is in the range 3.1-3.6 MeV/fm$^3$, while $\kappa$ is within 1.8-1.9.
One can see in Fig.~\ref{fig.7} that at $A=78$ the symmetry energy $S(A)$ as
well as its volume and surface components, $S^V$ and $S^S$ and pressure, $p_0$
reach minimum values while $\kappa$ has a very weak maximum ($\kappa$=1.9).

The smaller values of the symmetry energies of NM observed in Fig.~\ref{fig.5}
at the range of densities relevant in the integration in Eq.~(\ref{eq:27}) lead
to smaller values for the NSE in the case of Bonn B and Bonn CD potentials from
\cite{Machleidt_ANP1989,Machleidt2001} in comparison with the Brueckner
potential in the Ni chain. At the same time, the almost identical density
dependence of $p_{0}^{NM}(\rho)$ of both realistic potentials (see
Fig.~\ref{fig.2}) produce similar values of $p_{0}$ for the Ni isotopes being
larger than the Brueckner ones.

Here we note that the observed peaks in the NSE and its volume and surface
components at $A=78$ (see Fig.~\ref{fig.7}) take place for all choices of Skyrme
interaction parametrizations (SLy4, Sk3, and SGII). They are more pronounced for
the choice of the Brueckner energy density functional and they are somewhat
smoothed out and less pronounced for Bonn B and Bonn CD one-boson-exchange
potentials. We attribute this peak to the abrupt nuclear density change that is
characteristic for double-magic nuclei, such as $^{78}$Ni. We note that these
results concern all three Skyrme forces used (SLy4, Sk3, and SGII).

Similar peaks in the symmetry energy and its surface and volume components as
functions of the mass number are predicted at the double-magic nucleus with
$A=132$ in Sn (Fig.~\ref{fig.8}) (and additional peak at the semi-magic nucleus
with mass number $A=140$ observed for the volume contribution to the NSE) and no
pronounced peak at the double-magic nucleus with $A=208$ in Pb
(Fig.~\ref{fig.9}) for all three choices of the parametrization of the Skyrme
interaction. These peaks are more pronounced for the Brueckner density
functional and much less pronounced for the employed Bonn B and Bonn CD
potentials \cite{Machleidt_ANP1989,Machleidt2001}. In addition, the conclusions
drawn already on the magnitude of the NSE $S$ and pressure $p_{0}$ for Ni
isotopes using the three potentials and three Skyrme forces are valid also for
the cases of Sn and Pb isotopes.

Here we would like to note that the observed maxima and minima of the considered
quantities on Figs.~\ref{fig.7}-\ref{fig.9} had been found and discussed in our
previous works \cite{Gaidarov2011,Gaidarov2012,Antonov2017,Antonov2018}
(so-called there ``kinks''), studying the density dependence of the NSE for Ni,
Sn, and Pb isotopes (see, e.g., details in Ref.~\cite{Gaidarov2012}).They are
related to the shell effects which are important at zero temperature. As
mentioned above, the observed behavior of $S$, $S^V$, $S^S$, and $\kappa$ can be
attributed to the profiles of the density distributions, particularly in the
surface region. We mention that even the small differences between the densities
of the double-magic nuclei $^{78}$Ni, $^{132}$Sn (and also of the semi-magic
$^{140}$Sn where the neutron shell $2f_{7/2}$ is closed), as well as $^{208}$Pb,
and the neighbor nuclei lead through their derivative to larger differences of
the weight function $|{\cal F}(x)|^{2}$ [see Eq.~(\ref{eq:25})]. The latter is
one of the main ingredients in the calculations of the quantities studied in our
work [see Eqs.~(\ref{eq:27}), (\ref{eq:28}), and (\ref{eq:32})].

Here we would like to note that in some cases the observed peculiarities (maxima
and minima, the so called “kinks”) in Figs.~\ref{fig.7} and \ref{fig.8} are
different, from one side, when the Bonn B and Bonn CD potentials are used and,
from other side, in the case of the Brueckner EDF. These are, for example, the
cases of $S^S$, $\kappa$ and $p_0$ (with the SLy4 force) for Ni isotopes and
$S$, $S^S$, $\kappa$ and $p_0$ in Fig.~\ref{fig.8} for Sn isotopes. Thus, within
our method, in addition to the profiles of the density distributions, the reason
for the mentioned differences is also related to the model used to calculate the
nuclear matter properties that enter the calculations of the corresponding
finite nuclear properties under study.

The values of the symmetry energy, of its volume and surface components and
their ratio obtained in our work and shown above are, in general, compatible
with those quoted in the existing literature (see, e.g.,
\cite{Dan2003,Dan2004,Danielewicz,Dan2014,Han88}). The latter are, for example,
the empirical value of the symmetry energy $ 30\pm 4 $ MeV given in
Refs.~\cite{Han88,Nik2011}, the values of the volume symmetry energy (between 27
and 33.7 MeV) \cite{Dan2003,Dan2004,Danielewicz,Dan2014} and the surface
symmetry energy (between 9 and 12 MeV) \cite{Danielewicz}. In addition, the
published values of $\kappa$ extracted from nuclear properties, such as the
isobaric analog states and skins \cite{Dan2004} and masses and skins
\cite{Dan2003}, being presented in \cite{Diep2007}, are $2.0 \le \kappa \le
2.8$. Another range of values of $\kappa$ ($ 1.6 \le \kappa \le 2.0 $) is also
given in Ref.~\cite{Diep2007}.

Considering in more details the comparison of our results for the NSE and its
components with the existent data mentioned above we note that the main
differences exist in the following cases: i) for the surface component of NSE
$S^S$ between 14.1 and 14.6 MeV when the Bonn CD potential is used and between
14.9 and 16.0 in the case of the Bonn B potential, and ii) for the NSE $S$
(between 24.4 and 24.8 MeV) when the Bonn B potential is used.

The values of the pressure obtained in our work for the three chains of isotopes
(Ni, Sn, and Pb) using Bonn B and Bonn CD potentials from
\cite{Machleidt_ANP1989,Machleidt2001} are around 3 MeV/fm$^{3}$. This value is
close to the upper limit of the range of 1.95-2.95 MeV/fm$^{3}$ of theoretical
estimates in other Brueckner-Hartree-Fock calculations \cite{Diep2003} with
various versions of Argonne $V_{18}$ potentials and is compatible with the value
$p_0=2.3 \pm 0.8$ MeV/fm$^{3}$ extracted from measurements of the strength of
the pygmy dipole resonance in Sn and Pb isotopes \cite{Klimkiewicz_et2007}.
However, the value of $p_0$ obtained in our work when the Brueckner EDF is used
is roughly twice less than those obtained when we use Bonn B and Bonn CD
potentials \cite{Machleidt_ANP1989,Machleidt2001} and than the already mentioned
values in Refs.~\cite{Diep2003} and \cite{Klimkiewicz_et2007}.

In addition, the values of the nuclear pressure $p_{0}$ obtained in our approach
with the same potentials for $^{208}$Pb nucleus agree well with the recent
results for the slope parameter $L$ between 44 and 58 MeV \cite{Sammarruca2019}
that correspond to $p_{0}$ between 2.35 and 3.09 MeV/fm$^{3}$. The latter values
obtained from most recent EOS based on state-of-the-art chiral $NN$ potentials
were extracted from the typical correlation between $L$ and the thickness of the
neutron skin in $^{208}$Pb (between approximately 0.14 and 0.16 fm).

\begin{figure*}[ht]
\centering
\includegraphics[width=0.6\textwidth]{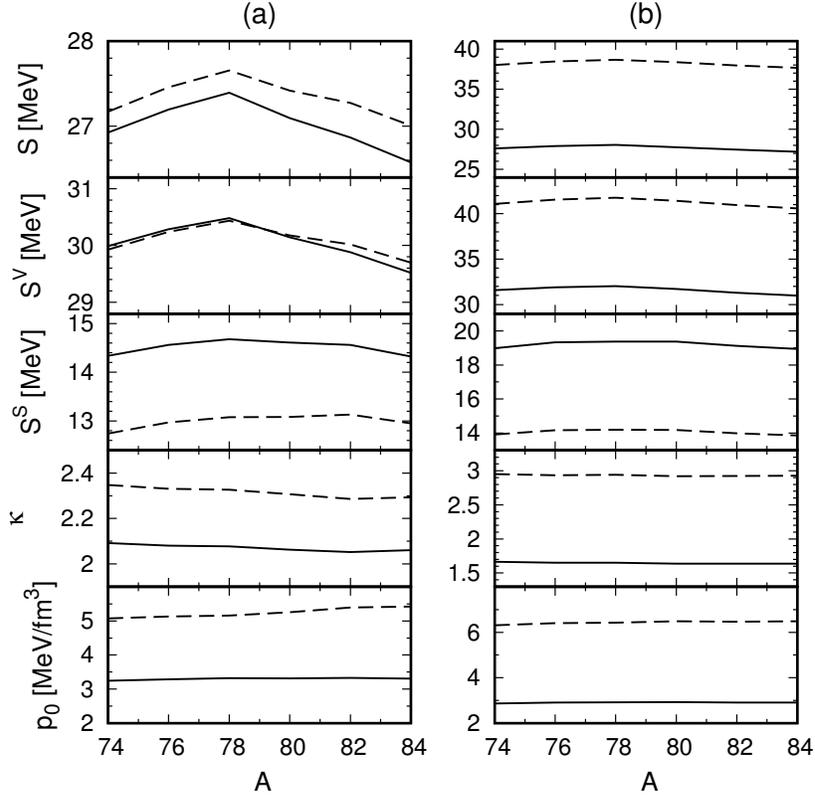}
\caption[]{The symmetry energy $S$, its volume $S^V$ and surface $S^S$
components, their ratio $\kappa$ and the pressure $p_0$ for Ni isotopes,
obtained by means of the CDFM and different nuclear matter calculations using
the Bonn B potential in the BHF method [54] (a) and the Bonn CD potential in the
in-medium T-matrix method [55] (b) . Solid (dashed) line corresponds to the
results in the case without (with) TBF contributions (see the text). The weight
function $|{\cal F}(x)|^{2}$ used in the calculations is obtained by means of
the densities derived within a self-consistent Skyrme-Hartree-Fock plus BCS
method with SLy4 interaction.
\label{fig.10}}
\end{figure*}

An extension to include TBF in finite systems, in particular in medium and heavy
neutron-rich nuclei, for analysis of EOS properties of these nuclei in the
framework of the BHF is an important task. In this case, the application of the
CDFM could be justified {\it a posteriori} as a good choice to calculate the
finite nuclei properties also when including three-body forces. Therefore, as a
first step in this direction we give, as an example, estimations for the
symmetry energy, its volume and surface contributions and their ratio, as well
as the pressure, in the case of Ni isotopes on top of BHF calculations by
including the corresponding self-consistent three-body forces. For this purpose
we employ, as examples, the considerations from Ref.~\cite{Wang2014} where the
BHF method with Bonn B potential is used, as well as from Ref.~\cite{Soma2008}
where the in-medium T-matrix approach and Bonn CD potential is used. The role of
the TBF in the case of the Bonn B potential [Fig.~\ref{fig.10}(a)] is to
increase the symmetry energy $S$ for Ni isotopes with $A=74$ - $84$ from 26.92 -
26.57 MeV to 27.17 - 27.01 MeV, to change $S^{V}$ from 30.00 - 29.52 MeV to
29.93 - 29.70 MeV, to decrease $S^{S}$ from 14.33 - 14.32 MeV to 12.74 - 12.95
MeV, as well as to increase $\kappa$ from 2.09 - 2.06 to 2.34 - 2.29 and $p_{0}$
from 3.34 - 3.30 to 5.08 - 5.42 MeV/fm$^{3}$. We note the existence of the
maxima of $S$ and $S^{V}$ for $^{78}$Ni in both cases of two- and three- body
forces and of $S^{S}$ in the case of two-body forces. The behaviour of the
calculated characteristics for the same Ni isotopes when TBF are included can be
seen in the case of the Bonn CD in Fig.~\ref{fig.10}(b). However, there is a
bigger increase of $S$ from 27.62-27.19 MeV to 38.02-37.65 MeV, an increase of
$S^{V}$ from 31.57-30.98 MeV to 41.09-40.58 MeV, a decrease of $S^{S}$ from
18.98-18.93 MeV to 13.92-13.86 MeV, an increase of $\kappa$ from 1.664-1.636 to
2.95-2.93, and strong one of $p_{0}$ from 2.86-2.911 MeV/fm$^{3}$ to 6.31-6.49
MeV/fm$^{3}$. One can see maxima for two- and three-body forces at $A=78$ in the
cases of $S$, $S^{V}$, and $\kappa$ and a maximum of $S^{S}$ only in the case of
TBF. The differences between the results obtained with the use of Bonn B and
Bonn CD potentials plus TBF can be attributed to different EOS of nuclear matter
(see Figs.~\ref{fig.1} and \ref{fig.3}) being stiffer for the Bonn CD potential.
In addition, we have to note the role of the weight function $|{\cal F}(x)|^{2}$
when averaging the NM properties for both potentials plus TBF to obtain the
considered quantities for the Ni isotopes in the CDFM scheme. Although the TBF
affect considerably the high-density behavior of the symmetry energy of nuclear
matter, their role on the symmetry energy and related quantities is not so large
at subsaturation densities relevant for finite nuclei. This is due to the fact
that the CDFM weight function $|{\cal F}(x)|^{2}$ in our approach has a bell
form (see Fig.~\ref{fig.6}), which is peaked at distances that correspond to a
density of around $\rho \approx 0.05-0.08$ fm$^{-3}$ and its width is within the
density range between 0.12 and 0.01 fm$^{-3}$. Namely in this region, where (as
can be seen from Figs.~\ref{fig.1} and \ref{fig.3}) the difference between the
values of $S^{NM}(\rho)$ in cases of both Bonn B and Bonn CD potentials when the
two- and three-body forces are used is not so large. The values of the weight
function are going to zero at densities $\rho > 0.12$ fm$^{-3}$ and $\rho <
0.01$ fm$^{-3}$ and that is why the differences between the values of
$S^{NM}(\rho)$ in the cases of two- and three-body forces for nuclear matter at
such densities are strongly reduced in finite nuclei.

\section{Conclusions \label{sec:5}}

In the present work we combine the non-relativistic Brueckner-Hartree-Fock
method with the realistic Bonn B and Bonn CD $NN$ potentials with the CDFM
within the self-consistent Skyrme-Hartree-Fock plus BCS approach to calculate
the nuclear symmetry energy, its volume and surface components and their ratio,
as well as the pressure for three isotopic chains of spherical nuclei (Ni, Sn,
and Pb). We find that the values of the NSE obtained in the BHF method are
consistent and within 1.0-2.0 MeV agreement with the values obtained with the
effective Brueckner potential. The calculated volume and surface components of
the NSE for the Ni, Sn, and Pb chains obtained from the realistic Bonn B and
Bonn CD two-body potentials are in a reasonable agreement with recently
published estimations and available experimental data.

The results of our calculations of the pressure using Bonn B and Bonn CD
potentials \cite{Machleidt_ANP1989,Machleidt2001} turn out to be close to the
upper limit of other BHF calculations using various versions of Argonne $V_{18}$
potentials and are comparable with the data from measurements of the strength of
the pygmy dipole resonance in Sn and Pb isotopes. At the same time, our values
of $p_{0}$ in the case when the Brueckner EDF is used are roughly twice less
than those obtained when Bonn B and Bonn CD potentials are used and the values
in Refs.~\cite{Diep2003,Klimkiewicz_et2007}.

Complementary, in the present work we estimate within our approach the effects
of three-body forces on the symmetry energy, its components and their ratio, as
a first step on the example of Ni isotopes using the CDFM with BHF approximation
with the Bonn B and Bonn CD potentials plus TBF \cite{Wang2014,Soma2008}. It is
shown that the bell form of the CDFM weight function $|{\cal F}(x)|^{2}$ that is
peaked around $\rho \approx 0.05$ - $0.08$ fm$^{-3}$ and whose values are going
to zero for $\rho >0.12$ and $\rho<0.01$ fm$^{-3}$, is the reason why the
difference between the symmetry energy (and related quantities) for nuclear
matter when the two- and three-body forces are used is much smaller in the case
of finite nuclei. The small effect when TBF are included is better observed in
the case of Bonn B three-body potential, where the values of the studied
quantities are closer to the ones calculated by adopting purely the two-body
forces. In our opinion, the role of microscopic three-body forces in the
proposed approach to study the surface properties of neutron-rich nuclei can be
clearly revealed in the future by applying, for instance, the latest version of
the Barcelona-Catania-Paris-Madrid nuclear energy density functional
(\cite{Sharma2015} and references therein), which is constructed upon the BHF
calculations in nuclear matter and is able to treat successfully medium-heavy
nuclei.
\begin{acknowledgments}
I.C.D would like to thank Professor Fred Myhrer and Professor Kuniharu Kubodera
for numerous discussions and valuable advice and Professors Ruprecht Machleidt
and Morten Hjorth-Jensen for their help in developing his computer code. I.C.D
also wishes to acknowledge partial financial support from the University of
Mount Olive Professional Development Fund. A.N.A, D.N.K., and M.K.G. are
grateful for the support of the Bulgarian Science Fund under Contract No.
KP-06-N38/1. P.S. acknowledges support from Ministerio de Ciencia, Innovaci\'on
y Universidades MCIU/AEI/FEDER,UE (Spain) under Contract No.
PGC2018-093636-B-I00.
\end{acknowledgments}

\end{document}